\documentclass[final,5p,times,twocolumn]{elsarticle}

\usepackage{amssymb}
\usepackage{amsmath}
\usepackage{multirow}
\usepackage[hidelinks]{hyperref}
\usepackage{xcolor}
\usepackage{lipsum}
\usepackage{tabularx}
\usepackage{setspace}
\definecolor{DeepRed}{RGB}{181 17 27}
\definecolor{DeepBlue}{RGB}{0 0 255}
\hypersetup{
	colorlinks,
	linkcolor={red!50!red},
	citecolor={blue!50!red}, %
	urlcolor={blue!80!red}
}
\usepackage{amsthm}

\usepackage[switch]{lineno}

\journal{Materials Today Physics}

\begin{document}
	
	\begin{frontmatter}
		
		\title{{\color{black}Analytical Modeling of Acoustic Exponential Materials and Physical Mechanism of Broadband Anti-Reflection}}
		
		\author[inst1,inst2]{Sichao Qu}
		\cortext[cor1]{Corresponding author(s).}
		\author[inst2]{Min Yang\corref{cor1}}
		\ead{min@metacoust.com}
		\author[inst2]{Tenglong Wu}
		\author[inst2]{Yunfei Xu}
		\author[inst1]{Nicholas Fang\corref{cor1}}
		\ead{nicxfang@hku.hk}
		\author[inst2]{Shuyu Chen}
		
		\affiliation[inst1]{organization={Department of Mechanical Engineering, The University of Hong Kong},
			addressline={Pokfulam Road}, 
			city={Hong Kong},
			country={China}}
		
		\affiliation[inst2]{organization={Acoustic Metamaterials Group Ltd.},
			addressline={Data Technology Hub, TKO Industrial Estate},
			city={Hong Kong},
			country={China}}
		
		\begin{abstract}
		{\color{black} Spatially} exponential distributions of material properties are ubiquitous in many natural and engineered systems, from the vertical distribution of the atmosphere to acoustic horns and anti-reflective coatings. These media seamlessly interface different impedances, enhancing wave transmission and reducing internal reflections. This work advances traditional transfer matrix theory by integrating analytical solutions for acoustic exponential materials, {\color{black} which possess exponential density and/or bulk modulus}, offering a more accurate predictive tool and {\color{black}revealing the physical mechanism of broadband anti-reflection} for sound propagation in such non-uniform materials. Leveraging this method, we designed an acoustic dipole array that {\color{black}effectively} mimics exponential mass distribution. Through experiments with precisely engineered micro-perforated plates, we demonstrate an ultra-low reflection rate of about 0.86\% across a wide frequency range from 420 Hz to 10,000 Hz. Our modified transfer matrix approach underpins the design of exponential materials, and our layering strategy for stacking acoustic dipoles suggests a pathway to more functional gradient acoustic metamaterials.
		\end{abstract}
		
		
		
		
		\begin{keyword}
			Exponential materials \sep  Generalized eigenmodes\sep Impedance matching\sep Broadband anti-reflection \sep Acoustic metamaterials
		\end{keyword}
		
	\end{frontmatter}
	
\section{Introduction}
\label{sec:introduction}
	
In the presence of a gravitational field, matter density often adopts an exponential gradient distribution, as exemplified by the atmospheric density variation with altitude \cite{berberan1997barometric,perna2002self} or the gradation observed in ground soil \cite{raspet1996surface, attenborough1985acoustical} and marine sediments \cite{tolstoy1963theory, tolstoy1965effect}. Remarkably, such media, despite their inhomogeneity, allow sound waves to propagate without reflective energy loss. This unique characteristic facilitates their use as intermediary layers to minimize reflection losses between media of different impedances, with the acoustic horn being a classic example. The genesis of using graded materials to mitigate interface reflections dates back to the seminal works of Lord Rayleigh \cite{strutt1887iv} and Fraunhofer \cite{fraunhofer1888joseph} in the 19th century. Since then, gradient materials have found extensive applications across optical and acoustic systems \cite{moore1980gradient,jin2019gradient}, including optical fibers \cite{kao1966dielectric}, light-emitting diodes \cite{lenef2018}, lenses \cite{yang2004focusing,wu2008graded}, and absorbers \cite{ding2012ultra,ghaffari2015design}. Unlike antireflection techniques reliant on Fabry-P\'erot resonances \cite{pozar2011microwave}, gradient materials offer the significant advantage of broadband operation, an attribute crucial for applications in solar cells \cite{aiken2000high,tu2023antireflection}, disordered media \cite{horodynski2022anti,kats2013vanadium}, aberrating layer \cite{shen2014anisotropic,yang2017impedance}, and cloaking devices \cite{chen2010acoustic}.

Two prevalent mathematical approaches are employed to model gradient media: the small reflection theory \cite{pozar2011microwave,pendergraft1993exact} and the transfer matrix method \cite{pozar2011microwave,jackson1999classical}. The small reflection theory offers a general solution for gradient materials, predicated on the assumption of weak reflection and is particularly useful for simulating multisection binomial and Chebyshev transformers \cite{collin1955theory,klopfenstein1956transmission} in TEM transmission lines \cite{pozar2011microwave,pendergraft1993exact} under the approximation of an almost constant refractive index. However, this theory falters when faced with the substantial refractive index variations typical of exponential materials (EMs). On the other hand, the transfer matrix method, utilising a discretisation strategy \cite{bethune1989optical}, is theoretically applicable to any gradient medium. Yet, its accuracy is compromised in media with large property gradients, necessitating a finer discretisation meshes and increased computational resources to minimise cumulative numerical errors  \cite{luce2022tmm}. Consequently, the development of more efficient numerical techniques for EMs is imperative for the advanced design and optimisation in practical applications.
	
\begin{figure*}[t!]
	\centering
	\includegraphics[width=0.8\textwidth]{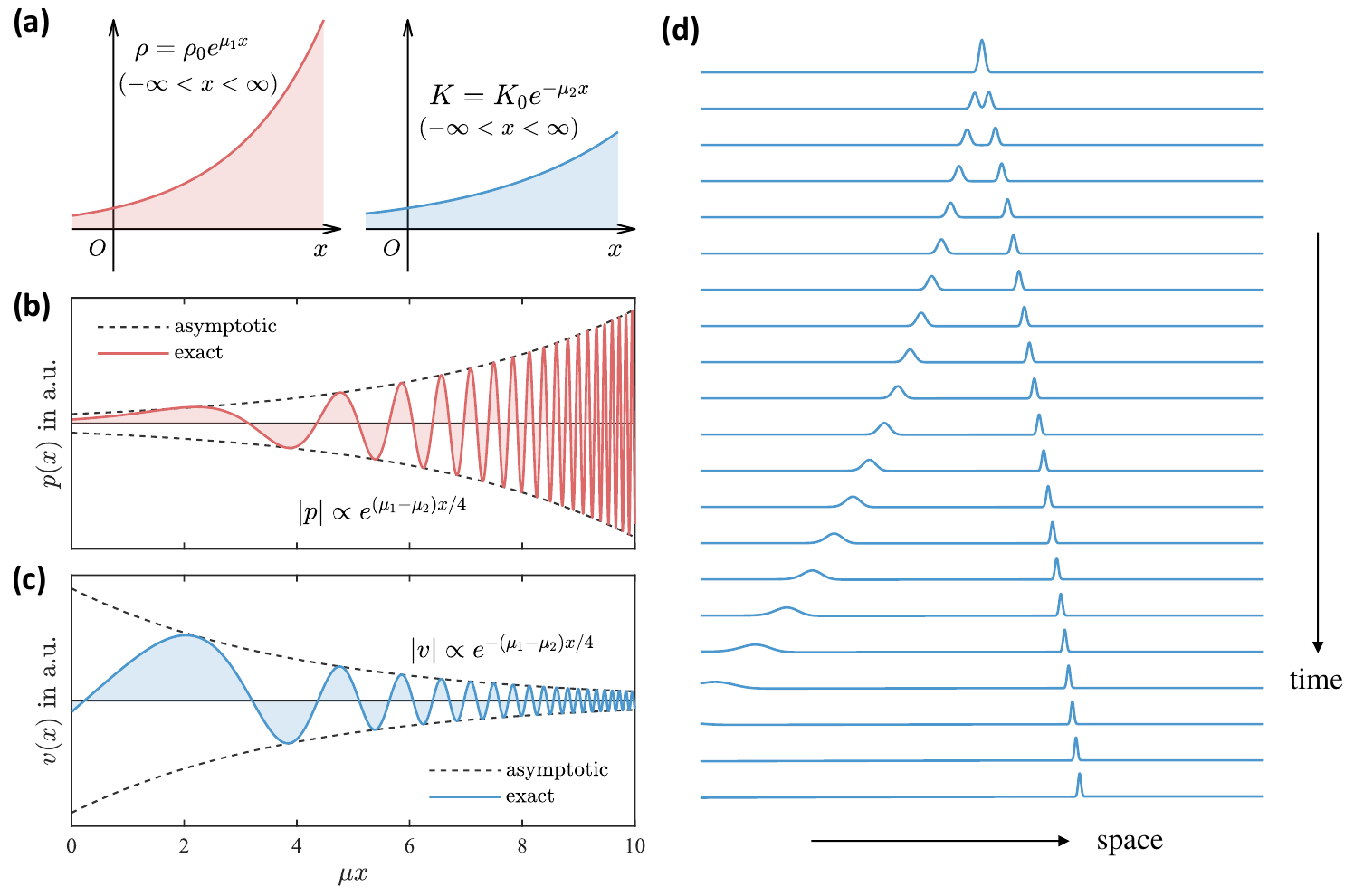}
	\caption{\label{fig:first} (a) The profiles of $\rho(x)$ and $\mathit{K}(x)$ for an ideal and infinitely long EM in the reduce region in Figure (\ref{fig:2}). Here we take the values when $\mu_1 = -2\mu_2/3$ for example. (b-c) The calculated eigenmodes ,i.e., pressure $ p $ and velocity $ v $, in this EM at dimensionless frequency $\xi_0 = f/f_0 = 4/5$. By plotting the results in Table~(\ref{tab:1}), the solid lines are real parts of exact solutions, while the dashed lines are the absolute values of asymptotic solutions. (d) The propagation of a Gaussian pulse in exponential materials with no reflection is given by Eq.~(\ref{eq:traveling_wave}), as shown in the numerical results.}
\end{figure*}

In this study, we innovate upon the classical transmission matrix method for uniform media by incorporating a plane wave eigen-solution specific to EMs. This refinement allows for a substantial reduction in the number of discrete meshes necessary for accurate simulations, thereby decreasing computational demands while enhancing precision \cite{anzengruber2012numerical,guo2014design}. The resulting method enables a more efficacious characterisation of material response properties, which is instrumental in optimising parameters and improving design processes for applications such as antireflection layers. Demonstrating the practical utility of our approach, we engineered a series of broadband acoustic dipoles using a sequence of non-resonant micro-perforated plates. These meta-layers were meticulously designed to yield a gradient in \emph{\color{black}effective mass density} that adheres to an exponential spatial distribution.  Here, the the prefix ``meta'' borrowed from metamaterials \cite{anzengruber2012numerical,guo2014design} emphasises the effective properties were from designed structures rather than chemical composition.  Experimental validation confirms the theoretical model, with the structure exhibiting exceptional broadband reflection suppression across an extensive frequency spectrum {\color{black}(an average reflectivity of 0.86\% and less than 10\% from 420 to 10,000 Hz)}, when backed by a rigid surface. These results highlight the considerable potential to advance the application of EMs in fields like antireflection technology.

\section{Exponential medium and its general solution \label{sec:general_solution}}

Sound propagates as a spread of density variation within a medium, a process characterized by localized interactions rather than long-range effects.  Therefore, the homogeneity in the medium is not expected to affect the local momentum equation or alter the constitutive properties at a given point:

\begin{align}
\label{eq:constitutive}
\left\{
\begin{array}{l}
	\displaystyle{\partial \bf v}/{\partial t}+\nabla p/\rho({\bf x})=0,\\
	\displaystyle{\partial p}/{\partial t}+\rho({\bf x}) c({\bf x})^2\nabla\cdot{\bf v}=0.
\end{array}\right.
\end{align}
Here, $p$ is the pressure modulation by sound, $\bf v$ is the particle velocity, $\rho$ is the local density and $c=\sqrt{(\partial p/\partial \rho)_S}$ is the local sound speed in adiabatic processes.  By eliminating $\bf v$, the Eq.~\eqref{eq:constitutive} gives the wave equation respect to $p$ for the non-uniform medium

\begin{equation}
\label{eq:wave_equation}
	\rho({\bf x})\nabla\cdot \left[\frac{1}{\rho({\bf x})}\nabla p({\bf x})\right]-\frac{1}{c({\bf x})^2}\frac{\partial^2}{\partial t^2}p({\bf x})=0.
\end{equation}

The problem reduces to one-dimensional if both $\rho$ and $c$ change exponentially along a single axis, $x$, and remain constant along the other two,

\begin{equation}
\left\{
\begin{array}{l}
	\rho({\bf x})=\rho_0\exp(\mu_1x),\\
	c({\bf x})=c_0\exp(-\mu x/2),
\end{array}\right.
\end{equation}
provided that the sound propagation under consideration is in the direction of the exponential variation.  Where, $\rho_0$ and $c_0$ are real-valued reference material constant defined at $x=0$. Also, we display the corresponding bulk modulus as

\begin{equation}\label{eq:Kx}
	K({\bf x}) = \rho c^2 = K_0 \exp(-\mu_2 x),
\end{equation}
where $\mu_2 = \mu -\mu_1$ followed from Eq.~\eqref{eq:constitutive}. The schematic diagram illustrating the exponential material properties is depicted in Figs.~(\ref{fig:first}a) and (\ref{fig:first}b). {\color{black}To find the solution to Eq.~\eqref{eq:wave_equation}, we introduce variable transformations for monochromatic sound\footnote{In this context, we assume a harmonic dependence of $ p $ as $\exp(-i\omega t)$, and we globally adopt ${\partial}/{\partial t}\to -i\omega t$.} with the angular frequency $\omega$,

\begin{equation}
	x\mapsto\xi\equiv\frac{2\omega}{\mu c_0}\exp(\mu x/2)\text{ and }
	p\mapsto\Phi\equiv \xi^{-\alpha_1}p,
\end{equation}
where $\alpha_1 = \mu_1/\mu$. So, Eq.~\eqref{eq:wave_equation} yields a Bessel equation:

\begin{equation}
	\xi^2\frac{\partial^2\Phi}{\partial\xi^2}+\xi\frac{\partial\Phi}{\partial\xi}+\left[\xi^2-\alpha_1^2\right]\Phi=0.
\end{equation}
The relevant general solutions for sound pressure can be expressed by Hankel functions of the first and second kinds (referring to {\color{DeepBlue}Supplementary Materials, Section S1} for details):

\begin{align}
	p(x, t)=e^{\frac{\mu_1x}{2}}\int_{0}^{\infty}&\left\{C_1(\omega)H_{\alpha_1}^{(1)}[\xi(\omega)]+\right.\nonumber\\
	&\left.+C_2(\omega)H_{\alpha_1}^{(2)}[\xi(\omega)]\right\}e^{-i\omega t}d\omega.
\label{eq:traveling_wave}
\end{align}
Here, $C_1$ and $C_2$ are determined by the initial condition of sound pressure's distribution at $t=0$, ensuring that the  requirement $p(x,0)=e^{\frac{\mu_1x}{2}}\int\left[C_1H_{\alpha_1}^{(1)}(\xi)+C_2H_{\alpha_1}^{(2)}(\xi)\right]d\omega$ is satisfied.}

\begin{table*}[t!]
	\centering
	\begin{tabular}{l|c|ccc}
		\hline\hline
		{}&\multicolumn{1}{c|}{Uniform Materials}&\multicolumn{3}{c}{Exponential materials (EMs)}\\
		Properties&\multicolumn{1}{c|}{$ \rho=\rho_0 , \mathit{K}=\mathit{K}_0$}&\multicolumn{3}{c}{$ \rho=\rho_0 e^{\mu_1 x}, \mathit{K}=\mathit{K}_0 e^{-\mu_2 x} $}\\
		\hline
		Solution type & Exact & Exact & \multicolumn{2}{|c}{ \text { Asymptotic }($ \xi \gg 1 $)}\\\hline
		$\begin{array}{l}
			\text{Pressure }p \\ \text{(Standing wave)}
		\end{array}$ & $\left[\begin{array}{c}\sin(k_0 x) \\ \cos(k_0 x)\end{array}\right]$ & $e^{\frac{\mu_1 x}{2}}\left[\begin{array}{l}J_{\alpha_1}(\xi) \\ Y_{\alpha_1}(\xi)\end{array}\right]$ & \multicolumn{2}{|c}{$\left(\frac{2}{\pi \xi_0}\right)^{\frac{1}{2}} e^{\frac{\left(\mu_1-\mu_2\right) x}{4}}\left[\begin{array}{c}\sin\left(\xi-\alpha_1 \frac{\pi}{2}-\frac{\pi}{4}\right)  \\ \cos\left(\xi-\alpha_1 \frac{\pi}{2}-\frac{\pi}{4}\right)\end{array}\right]$}\\\hline
		$\begin{array}{l}
			\text{Velocity }v \\ \text{(Standing wave)}
		\end{array}$ & $\frac{-i}{Z_0}\left[\begin{array}{c}\cos(k_0 x) \\ -\sin(k_0 x) \end{array}\right]$ & $\frac{-i e^{\frac{\mu_{2} x}{2}}}{Z_0}\left[\begin{array}{l}J_{-\alpha_2}(\xi) \\ Y_{-\alpha_2}(\xi)\end{array}\right]$ & \multicolumn{2}{|c}{$\left(\frac{2}{\pi \xi_0}\right)^{\frac{1}{2}} \frac{-i e^{-\frac{\left(\mu_1-\mu_2\right) x}{4}}}{Z_0}\left[\begin{array}{c}\sin\left(\xi+\alpha_2 \frac{\pi}{2}-\frac{\pi}{4}\right) \\ \cos\left(\xi+\alpha_2 \frac{\pi}{2}-\frac{\pi}{4}\right)\end{array}\right]$}\\\hline
		
		$\begin{array}{l}
			\text{Pressure }p \\ \text{(Traveling wave)}
		\end{array}$ & $\left[\begin{array}{c}e^{i k_0 x} \\ e^{-i k_0 x}\end{array}\right]$ & $e^{\frac{\mu_1 x}{2}}\left[\begin{array}{l}H_{\alpha_1}^{(1)}(\xi) \\ H_{\alpha_1}^{(2)}(\xi)\end{array}\right]$ & \multicolumn{2}{|c}{$\left(\frac{2}{\pi \xi_0}\right)^{\frac{1}{2}} e^{\frac{\left(\mu_1-\mu_2\right) x}{4}}\left[\begin{array}{c}e^{i\left(\xi-\alpha_1 \frac{\pi}{2}-\frac{\pi}{4}\right)} \\ e^{-i\left(\xi-\alpha_1 \frac{\pi}{2}-\frac{\pi}{4}\right)}\end{array}\right]$}\\\hline
		$\begin{array}{l}
			\text{Velocity }v \\ \text{(Traveling wave)}
		\end{array}$ & $\frac{1}{Z_0}\left[\begin{array}{c}e^{i k_0 x} \\ -e^{-i k_0 x}\end{array}\right]$ & $\frac{-i e^{\frac{\mu_{2} x}{2}}}{Z_0}\left[\begin{array}{l}H_{-\alpha_2}^{(1)}(\xi) \\ H_{-\alpha_2}^{(2)}(\xi)\end{array}\right]$ & \multicolumn{2}{|c}{$\left(\frac{2}{\pi \xi_0}\right)^{\frac{1}{2}} \frac{-i e^{-\frac{\left(\mu_1-\mu_2\right) x}{4}}}{Z_0}\left[\begin{array}{c}e^{i\left(\xi+\alpha_2 \frac{\pi}{2}-\frac{\pi}{4}\right)} \\ e^{-i\left(\xi+\alpha_2 \frac{\pi}{2}-\frac{\pi}{4}\right)}\end{array}\right]$}\\\hline
		\hline
	\end{tabular}
	\caption
	{Eigenmodes ($p$ and $v$) of uniform and exponential materials ($e^{-i\omega t}$ omitted). The displayed analytical solutions for EMs are the generalized version of those in uniform materials. In uniform materials, the wavenumber is defined as $k_0 = \omega /c_0$. At the high frequency limit when $\xi\gg 1$, the asymptotic eigenmodes \cite{stone2009mathematics} in EMs share the similar forms with plane-wave solution in uniform materials. However, the amplitude of $p$ (or $v$) is modulated by $e^{(\mu_1-\mu_2)x/4}$ (or $e^{-(\mu_1-\mu_2)x/4}$). The velocity fields are obtained by substituting $ p $ into the first line of Eq.~(\ref{eq:constitutive}) with the additionally defined $\alpha_2=1-\alpha_1$.}
	\label{tab:1}
\end{table*}

{\color{black}As an example, assuming the initial state comprises a Gaussian wave packet, $p(x,0)=\exp(-(ax)^2)$, the subsequent evolution of sound wave can be observed in Fig.~(\ref{fig:first}d) (numerical calculation steps are in {\color{DeepBlue}Supplementary Materials}, Section S2).  The wave traveling to the right is represented by the Hankel function $H_{\alpha_1}^{(1)}(\xi)$ in Eq.~\eqref{eq:traveling_wave} and manifests as $\sqrt{2/\pi\xi}e^{i(\xi-\frac{\alpha_1\pi}{2}-\frac{\pi}{4})}$ in the far field, whose profiles we have plotted in Figs.~(\ref{fig:first}b) and (\ref{fig:first}c).  Conversely, the wave moving to the left is denoted by $H_{\alpha_1}^{(2)}(\xi)$ in Eq.~\eqref{eq:traveling_wave} and exhibits behaviour $\sqrt{2/\pi\xi}e^{-i(\xi-\frac{\alpha_1\pi}{2}-\frac{\pi}{4})}$ at a significant distance from the source. The interpretation of standing and traveling plane waves is justified by the extremum of power flow (see a rigorous proof in {\color{DeepBlue}Supplementary Materials}, Section S3). Remarkably, despite the medium's non-uniformity, no reflections are observed for either of the propagating pulses, a distinctive characteristic of an EM.  

In Table~(\ref{tab:1}), we listed the plane-wave eigenmodes (in frequency domain) with respect to $ p $ and $v$, of both uniform materials and EMs for comparison. The detailed derivation is available in {\color{DeepBlue}Supplementary Materials}, Section S1.}

\begin{figure}[t!]
	\centering
	\includegraphics[width=0.5\textwidth]{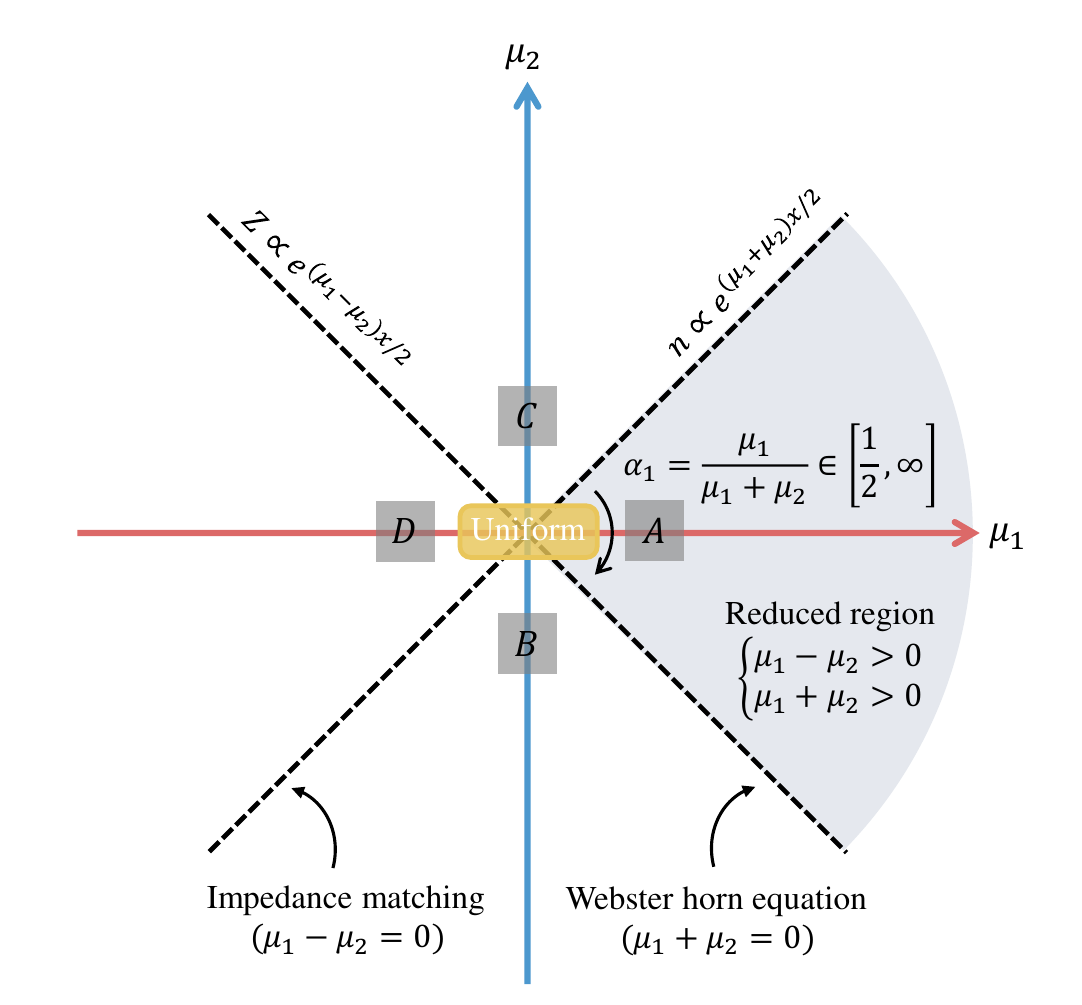}
	\caption{\label{fig:2} \color{black} The classification of EMs by the parameter space of $\mu_1$ and $\mu_2$, which are the measure of non-uniformity of exponential density and bulk modulus, respectively. In the center, $\mu_1=\mu_2=0$, which represents the space for uniform materials. We define EMs in other spaces, i.e., the farther away from the origin the greater the non-uniformity. By leveraging the duality, we have the following problems equivalent:  $A\Leftrightarrow C$ and $B\Leftrightarrow D$. Moreover, with parity symmetry, we have $A\Leftrightarrow D$ and $B\Leftrightarrow C$. Therefore, all possibilities of the solutions can be contained in the reduced region A.}
\end{figure}

\section{Classification of exponential materials}
\label{sec:classicfication_of_exponential_materials}

To lay the groundwork for advancing numerical methods applicable to exponential materials (EMs), it is instructive to first categorize and examine general solutions through the lens of symmetry. Eq.~\eqref{eq:constitutive} bears resemblance to the electric-magnetic duality (or Montonen–Olive duality) \cite{jackson1999classical} found in Maxwell's equations; by interchanging $p \leftrightarrow v$ and $\rho \leftrightarrow K^{-1}$, we observe that the equation retains its form, thereby exhibiting duality symmetry. This symmetry permits an interchange of $p$ and $v$ by simply swapping $\mu_1$ with $\mu_2$ and vice versa. Moreover, implementing a parity transformation $x \leftrightarrow -x$ is also mathematically tantamount to switching $\mu_1 \leftrightarrow -\mu_1$ and $\mu_2 \leftrightarrow -\mu_2$.

Nevertheless, existing references tend to name their materials using either gradient impedance \cite{pedersen1982impedance} or gradient index \cite{kim2013perfect}, while ignoring the other. Here, we show the necessity of considering both material properties to obtain a complete classification and to cover various EMs, as seen in the parameter space for classifying EMs. In Fig.~(\ref{fig:2}), we define the corresponding characteristic impedance $Z$ and refraction index $n$ as

\begin{equation} \label{eq:2}
	\left\{\begin{array}{c}
		Z(x)=Z_0 \exp \left[\frac{(\mu_1-\mu_2) x}{2}\right] \\
		n(x)=n_0 \exp \left[\frac{(\mu_1+\mu_2) x}{2}\right]
	\end{array}\right.,
\end{equation}
where $Z_0 = \sqrt{\rho_0 \mathit{K}_0}$ and $n_0 = 1$. Visualizing $\mu_1$ as the horizontal axis and $\mu_2$ as the vertical axis, we can represent all potential parameters of EMs on a two-dimensional plane. Given the symmetries described, the solution space of Eq.~\eqref{eq:wave_equation} can be effectively narrowed down to the shaded region A in Fig.~(\ref{fig:2}), characterized by $\mu_1 - \mu_2 > 0$ and $\mu_1 + \mu_2 > 0$. For the purposes of our research, we will confine our analysis to this region without compromising generality. Along the boundary where $\mu_1 - \mu_2 = 0$, the material exhibits a constant acoustic impedance, $Z = \rho_0 c_0 \exp[(\mu_1 - \mu_2)x/2]$, while along the boundary where $\mu_1 + \mu_2 = 0$, the speed of sound within the material remains invariant, which is the case of sound in a horn characterized by Webster equation \cite{martin2004webster}.  The point where $\mu_1 = \mu_2 = 0$ corresponds to a uniform medium, as depicted at the origin in Fig.~(\ref{fig:2}).

	\section{Modified transfer matrix method}
	\subsection{Model setup}
\begin{figure*}[t!]
		\centering
		\includegraphics[width=0.85\textwidth]{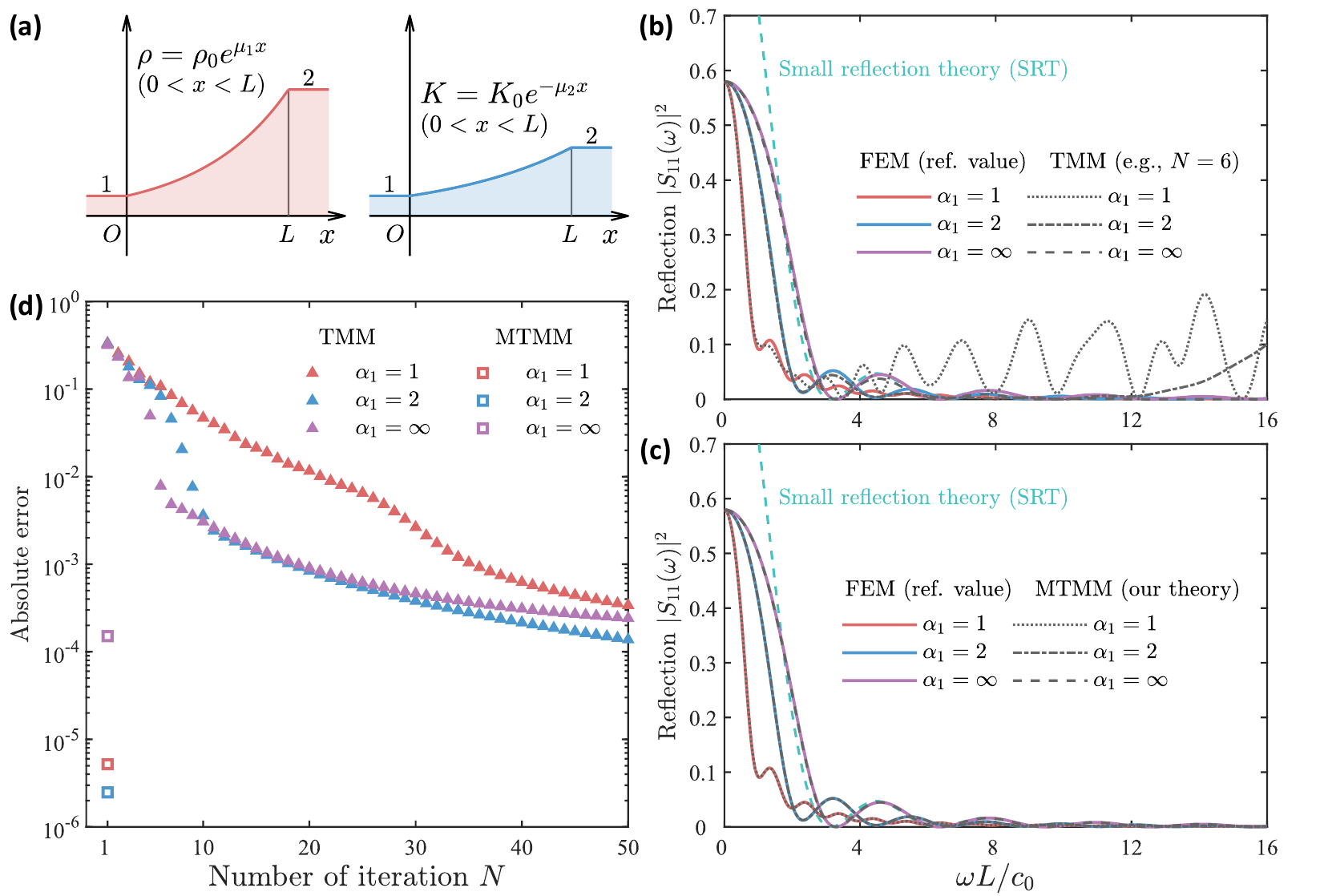}
		\caption{\label{fig:3} (a) The impedance transformer setup. The left and right sides are uniform materials with distinct material properties, which are connected with an EM. {\color{black}The combination of $ (\mu_1,\mu_2) $ can be arbitrary, reflecting the universality of our model.} (b) The reflection $|S_{11}|^2$ predicted by traditional piece TMM ($N=6$), compared with that by FEM as reference value. (c) The reflection $|S_{11}|^2$ predicted by MTMM and FEM. The green dashed lines in (b) and (c) are the same data given by SRT. (d) The absolute error of TMM and MTMM, plotted as function of iteration number $N$. The coefficient $\alpha_1$ is adjusted for checking the generality and accuracy of different models.}
	\end{figure*}

	As shown in Fig.~(\ref{fig:3}a), the basic model setup involves considering two uniform materials with distinct material properties. The EM is embedded in between as an impedance transformer with thickness $L$, and the material properties are ensured to be continuously connected at the interfaces. In this way, the density and bulk modulus are defined as
	
	\begin{align}	
		\rho(x) &=  \left\{\begin{array}{cr}
			\rho_0 & x\leq0\\
			\rho_0 \exp \left(\mu_1 x\right)& 0< x < L \\
			\rho_0 \exp \left(\mu_1 L\right)& x\geq L
		\end{array}\right., \label{eq:mat_dis1} 
	\end{align}
	and
	\begin{align}	
		\mathit{K}(x) &=  \left\{\begin{array}{cr}
			\mathit{K}_0 & x\leq 0\\
			\mathit{K}_0 \exp \left(-\mu_2 x\right)& 0<x<L \\
			\mathit{K}_0 \exp \left(-\mu_2 L\right)& x\geq L
		\end{array}\right..\label{eq:mat_dis2}
	\end{align}
	As an additional constraint, we assume that the impedance contrast is a constant for all EMs with different testing $\alpha_1$, which ensures that
	
	\begin{equation}\label{eq:ratio}
		\frac{Z_L}{Z_0} =\exp \left[\frac{(\mu_1-\mu_2) L}{2}\right]= 7.4,
	\end{equation}
	which we will use as a demonstration value for all followed numerical results.

	\begin{table*}[t!]
		\begin{center}
			\begin{tabular}{c|c|c|c}
				\hline\hline
				\multicolumn{2}{c|}{Methods}& ABCD-matrix &Reflection ($ S_{11} $)  \\\hline
				Our theory &MTMM&   $ \begin{array}{c} \frac{\pi \xi_0}{2} \left(\begin{array}{cc} \sqrt{\frac{Z_L}{Z_0}}  e^{\frac{\mu_2 L}{2}} \mathcal{F}\left(\alpha_1,-\alpha_2, \xi_0, \xi_L\right) & -i \sqrt{\frac{Z_0}{Z_L}}   e^{\frac{\mu_1 L}{2}} \mathcal{F}\left(\alpha_1, \alpha_1, \xi_0, \xi_L\right) \\ -i\sqrt{\frac{Z_L}{Z_0}}  e^{\frac{\mu_2 L}{2}} \mathcal{F}\left(-\alpha_2,-\alpha_2, \xi_0, \xi_L\right)  & \sqrt{\frac{Z_0}{Z_L}}  e^{\frac{\mu_1 L}{2}} \mathcal{F}\left(\alpha_1,-\alpha_2, \xi_L, \xi_0\right) \end{array}\right) \\
					\text{where }\mathcal{F}\left(v_1, v_2, x_1, x_2\right)=J_{v_1}\left(x_1\right)Y_{v_2}\left(x_2\right)-J_{v_2}\left(x_2\right)Y_{v_1}\left(x_1\right)
				\end{array}
				$ & Eq.~(\ref{eq:s11_mtmm}) \\ \hline
				\multirow{2}{*}{Textbook}&SRT& \multicolumn{1}{c|}{Not applicable} & Eq.~(\ref{eq:SRT}) \\\cline{2-4}
				&TMM & $ \left(\begin{array}{cc}
					\frac{1}{\sqrt{Z_0}}& 0 \\
					0 & \sqrt{Z_0}
				\end{array}\right) \prod_{n=1}^{N} \left(\begin{array}{cc}\cos(k_n \frac{L}{N}) & -i Z_n \sin(k_n \frac{L}{N}) \\ -\frac{i} {Z_n}\sin(k_n \frac{L}{N}) &  \cos(k_n \frac{L}{N})\end{array}\right) \left(\begin{array}{cc}
					\sqrt{Z_L}& 0 \\
					0 & \frac{1}{\sqrt{Z_L}}
				\end{array}\right) $& Eq.~(\ref{eq:s11_mtmm})\\\cline{1-3}
				\hline\hline
		\end{tabular}\end{center}
		\caption
		{The theories for modeling EMs. For modified transfer matrix method (MTMM), it is defined that $\xi_0 = \xi|_{x=0}$ and $\xi_L = \xi|_{x=L}$. For small reflection theory (SRT), there is no matrix element involved. For traditional transfer matrix method (TMM), the ABCD-matrix is obtained by multiplying $ N $ sub-matrices, with the mesh size of $L/N$. In the $n_\mathrm{th}$ layer, the wavenumber $k_n =  \omega \sqrt{\rho_n/\mathit{K}_n}$ and the characteristic impedance $Z_n = \sqrt{\rho_n \mathit{K}_n}$, where $\rho_n = \rho_0 \exp(\mu_1 nL/N)$ and $\mathit{K}_n = \mathit{K}_0 \exp(-\mu_2 nL/N)$.}
		\label{tab:2}
	\end{table*}
	\subsection{Elements of modified transfer matrix}
	Since the analytical solutions obtained earlier in Table (\ref{tab:1}) are accurate and universal, we now establish a modified transfer matrix method (MTMM) by taking advantage of the generalized eigenmodes in EMs, to predict the scattering parameters. The uniform material on the left (or right) side is associated with Port 1 (or Port 2). The ABCD-matrix \cite{pozar2011microwave} of the 2-port system relates the fields at different ports in following way
	
	\begin{equation} \label{eq:transfer_relation}
		\left(\begin{array}{c}\frac{\left.p\right|_{x=0}}{\sqrt{Z_0}} \\ \left.v\right|_{x=0} \sqrt{Z_0}\end{array}\right)=\left(\begin{array}{ll}A & B \\ C & D\end{array}\right)\left(\begin{array}{c}\frac{\left.p\right|_{x=L}}{\sqrt{Z_L}} \\ \left.v\right|_{x=L} \sqrt{Z_L}\end{array}\right),
	\end{equation}
	where $A,B,C,D$ are dimensionless matrix elements that follows the convention of generalized scattering matrix formalism \cite{pozar2011microwave,jalas2013and}.
	If we apply the incident excitation from left and right sides respectively, the linear superposition of $p$ and $v$ can be written as
	
	\begin{equation} \label{eq:4x4}
		\left\{\begin{array}{c}
			\left(\begin{array}{l}
				1+S_{11} \\
				1-S_{11}
			\end{array}\right)=\left(\begin{array}{ll}
				A & B \\
				C & D
			\end{array}\right)\left(\begin{array}{l}
				S_{21} \\
				S_{21}
			\end{array}\right)
			\\
			\left(\begin{array}{c}
				S_{12} \\
				-S_{12}
			\end{array}\right)=\left(\begin{array}{ll}
				A & B \\
				C & D
			\end{array}\right)\left(\begin{array}{c}
				1+S_{22} \\
				-\left(1-S_{22}\right)
			\end{array}\right) 
		\end{array}\right.,
	\end{equation}
	where $S_{ij}$ are the scattering matrix elements whose absolute values squared represents the energy ratio from Port $i$ to Port $j$. If $i=j$, $|S_{ii}|^2$ denotes the reflected energy ratio at Port $i$. By treating $ A,B,C,D $ as the known elements, we can attain the scattering matrix
	
	\begin{equation}\label{eq:s11_mtmm}
		\left\{\begin{array}{l}
			S_{11}  =(A+B-C-D)/(A+B+C+D)  \\ 
			S_{12}  =2(A D-B C)/(A+B+C+D) \\
			S_{21}  =2/(A+B+C+D)\\
			S_{22}  = (-A+B-C+D)/(A+B+C+D)
		\end{array}\right.,
	\end{equation}
	where the analytical forms of $A,B,C,D$ elements of MTMM are given in Table (\ref{tab:2}), whose derivation has been placed in {\color{DeepBlue}Supplementary Materials}, Section S4.
	
	\subsection{S-matrix analysis}
	Because our system is time-invariant, linear and with scalar material properties, the \textit{reciprocity} holds \cite{jalas2013and}, which ensures that $S_{12} = S_{21}$ [thus $AD-BC=1$, according to Eq.~(\ref{eq:s11_mtmm})]. Therefore, the reciprocity can guarantee symmetrical transmission, i.e.,
	
	\begin{equation}\label{eq:sym_trans}
		|S_{12}|^2 = |S_{21}|^2.
	\end{equation}
	Since the embedded EM is lossless, i.e., $\mathrm{Im}(\rho)=0$, $\mathrm{Im}(\mathit{K})=0$, the absorption inside the EM should be zero, thus $|S_{12}|^2 +|S_{11}|^2 = 1$ or $|S_{21}|^2 +|S_{22}|^2 = 1$. With the consideration of Eq.~(\ref{eq:sym_trans}), we can conclude that
	
	\begin{equation}\label{eq:sym_refle}
		|S_{11}|^2 = |S_{22}|^2,
	\end{equation}
	although $S_{11}\neq S_{22}$ due to the phase difference.
	It should be noted that Eq.~(\ref{eq:sym_trans}) holds for all frequencies regardless of the types of the EM, while Eq.~(\ref{eq:sym_refle}) is valid only  when the EM is lossless (the case we focus on here). Symmetrical reflected energy allows us to focus on the case where the incident excitation is from only Port 1, without losing the generality.

	\subsection{Our model vs textbook theories}
	To compare the proposed MTMM with the two representative theories in textbooks, we treat the reflection spectra obtained from finite element method (FEM) model as the reference values. The commercial software, COMSOL Multiphysics, was utilized to implement FEM calculation throughout the paper. To ensure the correctness of FEM, the adopted mesh size was sufficiently small compared to the wavelength $\lambda$. The related results are displayed by the solid lines in Figs.~(\ref{fig:3}b) and (\ref{fig:3}c), for EMs with different $\alpha_1$. Remarkably, the corresponding data given by the first line of Eq.~(\ref{eq:s11_mtmm}) [see MTMM results in Fig.~(\ref{fig:3}c)] match the reference values for all $\alpha_1$ at all frequencies, which shows the generality and accuracy of our theory.
	
	For predicting reflection from impedance-varying materials, small reflection theory (SRT) is a widely-used lightweight method \cite{pozar2011microwave}. Its assumptions are that the reflection at each layer is a small quantity, and that wave speed $c_0$ (or $n_0$) is constant, i.e., $\alpha_1=\infty$ in our definition. So, analytical solutions may be obtained for special impedance distribution $ Z(x) $. For example, for the EMs we considered, the overall reflection coefficient has the following form \cite{pozar2011microwave, pendergraft1993exact}
	
	\begin{align}\hspace{-0.18cm} \color{black}
		S_{11} =\int_{0}^{L} \frac{e^{2 i k_0 x}}{2}\frac{\mathrm{d}}{\mathrm{d}x}\ln\left( \frac{Z(x)}{Z_0}\right){\mathrm{d}x}= \ln\left(\frac{Z_L}{Z_0} \right) \frac{e^{i k_0 L}\sin(k_0 L)}{2 k_0 L},
		\label{eq:SRT}
	\end{align}
	where $k_0 = \omega/c_0$. As shown by the green dashed lines in Figs.~(\ref{fig:3}b) and (\ref{fig:3}c), the data by SRT coincide with reference values only for $\alpha_1\to\infty$ and when the reflection is lower than 0.05. This is consistent with the prescribed assumptions of SRT.
	
	The second textbook method is traditional transfer matrix method (TMM), which requires piecewise discretization of the investigated materials. To yield $S_{11}$ coefficient, TMM also follows the similar procedure of obtaining ABCD-matrix, using the same formula, i.e., Eq.~(\ref{eq:s11_mtmm}). However, the overall ABCD-matrix is generated from the one-by-one multiplication of $N$ sub-matrices. See the mathematical details of TMM in {\color{DeepBlue}Supplementary Materials}, Section S4. As shown in Fig.~(\ref{fig:3}b), when $N=6$, reflection spectra by TMM is accurate at low frequencies ($\omega L/c_0 <4$), while for high frequencies ($\omega L/c_0 >4$), TMM suffers from the discretization approximation. If $N=50$, the error becomes low enough but the computation time is surged. Compared with SRT, TMM is general for all $\alpha_1$ but the accuracy is not ensured if $N$ is not sufficiently large.
	
	From this perspective, SRT and TMM are neither general nor accurate. TMM can be general with cost of iteration times $N$. In Fig.~(\ref{fig:3}d), we compare the absolute errors by TMM and MTMM in our case of EMs, respectively. Absolute errors denotes the frequency-averaged difference between the target theory and FEM. So, MTMM outperforms TMM because only one-step calculation is enough and the error of MTMM is smaller than that of TMM with even $N=50$.
	
	By equating left-side terms of Eq.~(\ref{eq:transfer_relation}) and the first line of Eq.~(\ref{eq:4x4}), the reflection can be related with the specific impedance $Z_s$
	
	\begin{equation}\label{eq:imp_mat}
		S_{11} = \frac{Z_s - Z_0}{Z_s + Z_0},
	\end{equation}
	where $Z_s = (p/v)|_{x=0}$. The mechanism-level understanding impedance matching and anti-reflection performance of EMs in Fig.~(\ref{fig:3}) requires analytical analysis on $Z_s$, which will be addressed next.
	
\begin{figure*}[t!]
		\centering
		\includegraphics[width=0.8\textwidth]{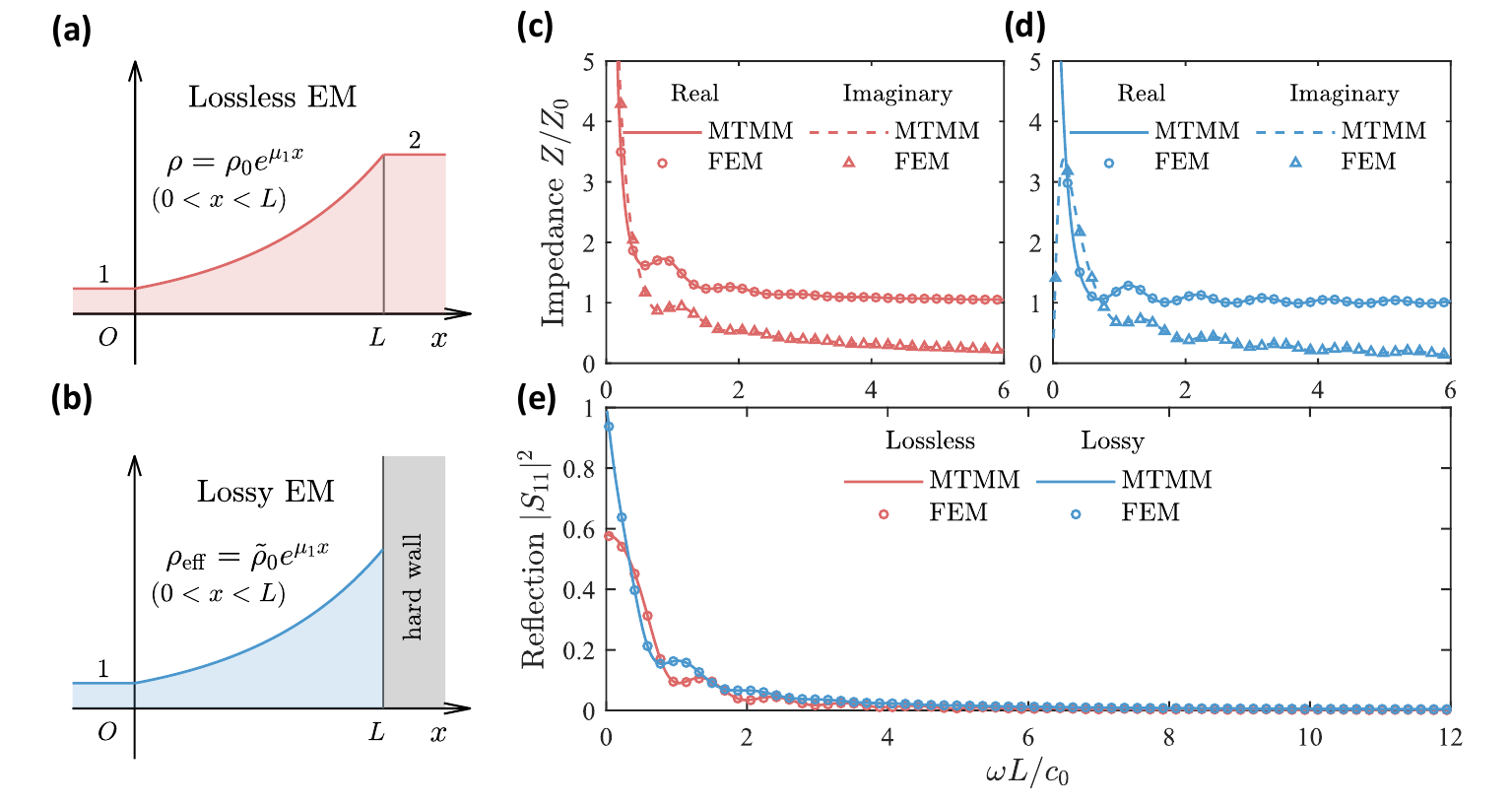}
		\caption{\label{fig:4} Lossless and lossy EMs with exponential density ($\alpha_1=1$). (a) The setup for a lossless EM backed by a leaky impedance boundary $Z(L) = Z_0 \exp \left[\frac{\mu_1 L}{2}\right]$. (b) The setup for a lossy EM with complex density $\tilde{\rho}_0 e^{\mu_1 x}$, backed by hard wall (i.e., $Z(L) \to \infty$). (c-d) The impedance spectra of lossless and lossy EMs in the setup of (a) and (b). (e) The reflection spectra of lossless and lossy EMs, given by Eq.~(\ref{eq:s11_mtmm}) and Eq.~(\ref{eq:S11_analy_hard}) respectively. The MTMM-predicted data agree well with those by FEM, showing the validity of our theory in lossy case.}
	\end{figure*}

	{\color{black}
	\section{Physical mechanism of broadband impedance matching}
	
	{\color{black}From Fig. (\ref{fig:3}c), it can be seen that in all cases, low-frequency reflection is still significant, while from intermediate to high frequency bands, the reflection tends to disappear, indicating broadband impedance matching. It is necessary to first understand how impedance behaves when there is a low-frequency mismatch in order to comprehend how it changes as the frequency increases.}

	\subsection{Low frequency behavior} 
	
	If $ \omega L /c_0\to 0 $, the ABCD-matrix of MTMM in Table~(\ref{tab:2}) becomes a diagonal matrix $ \left(\begin{array}{cc}
		\sqrt{Z_L/Z_0}& 0 \\
		0 & \sqrt{Z_0/Z_L}
	\end{array}\right) $, which has been proved in the {\color{DeepBlue}Supplementary Materials}, Section S4. According to Eq.~(\ref{eq:transfer_relation}), we conclude
	
	\begin{equation}\label{eq:limit1}
		\lim_{\omega\to0}Z_s =  Z_L,
	\end{equation}
	thus yielding the reflection $ S_{11} = (Z_L-Z_0)/(Z_L+Z_0) =0.58$. So, we can interpret it as the ‘bypassing' effect in EMs with our analytical model. From the perspective of impedance transfer, this means that the impedance at $ x=L $ is transmitted unchanged to $ x=0 $. {\color{black}In other words, the long wavelength wave can ignore impedance transition, and the reflection coefficient approaches the case without an anti-reflection layer, i.e., step impedance change.}
	
	\subsection{High frequency behavior}
	As shown in Fig.~(\ref{fig:3}c), if $\omega L /c_0 \to\infty$, we observe a near-zero reflection in all EMs. Now we explain why. The ABCD-matrix of MTMM in Table~(\ref{tab:2}) at the high frequency limit ($\omega\to\infty$) has the asymptotic values $A = D =1$ and $B=C=0$ (i.e. identity matrix). The related proof can also be found in the {\color{DeepBlue}Supplementary Materials}, Section S4. By adopting these values into Eq.~(\ref{eq:transfer_relation}), we have
	
	\begin{equation}\label{eq:limit2}
		\lim_{\omega\to\infty} Z_s = Z_0\frac{\left({p}/{v}\right) |_{x=L}}{Z_L}=Z_0.
	\end{equation}
	The last equal in Eq.~(\ref{eq:limit2}) is because the continuity impedance ensured by Eq.~(\ref{eq:ratio}), i.e., $\left({p}/{v}\right) |_{x=L}=Z(L)=Z_0e^{(\mu_1-\mu_2)/2}$. This means that any end impedance will be transformed into the impedance that matches Port 1, i.e., analytical evidence of the excellent impedance-matching feature of EMs. 
	
	\subsection{Intermediate frequency behavior}
	For the intermediate frequency range, we can see that impedance matching condition of Eq.~(\ref{eq:limit2}) is still a good approximation. This also explains why gradually varying media always have excellent anti-reflection properties \cite{jin2019gradient}. At high frequencies, the short wavelength makes it difficult to detect the non-uniformity of the material. As shown in Fig.~(\ref{fig:3}c), it can be seen that the anti-reflection properties can be maintained over a wide frequency range, depending on how small of $\alpha_1$ can be achieved. For instance, if we define $ f_1 $ as the threshold at which the reflection is below 0.1, then $ f_1 $ is lower and the anti-reflection effect is wider in bandwidth as $ \alpha_1 $ approaches $ 1/2 $. In the following content, we will focus on the case where $\alpha_1 = 1$, which is an achievable target in airborne acoustics. Furthermore, we will also examine the case with lossy material properties and its impact on impedance matching.
	
	\section{Lossless and lossy EMs}
	We first consider a lossless EM with $\alpha_1=1$ with the following settings:
	
	\begin{align}
		&\left\{\begin{array}{l}
			\rho(x) = \rho_0 e^{\mu_1 x}\\
			\mathit{K}(x) = \mathit{K}_0 
		\end{array}\right., \label{eq:lossless1} \\
		& \mathrm{with\;}  Z(L)=\frac{p}{v}\Big|_{x=L} = Z_0 e^ {\mu_1 L/2}, \label{eq:lossless2}
	\end{align}
	as shown in Fig.~(\ref{fig:4}a). Here, we set the impedance boundary as Eq.~(\ref{eq:lossless2}), following the same definition in Fig.~(\ref{fig:3}a). The product $\mu_1 L$ should be determined by Eq.~(\ref{eq:ratio}), i.e., $\mu_1 L = 4$. Here, Eq.~(\ref{eq:lossless2}) is regarded as leaky backing.
	
	We already know from Fig.~(\ref{fig:3}d) about the excellent anti-reflection properties of such EM, and the same curve is displayed by blue data in Fig.~(\ref{fig:4}e). Next, we consider the other lossy EM, with the settings:
	
	\begin{align}
		&\left\{\begin{array}{l}
			\rho(x) = \tilde{\rho}_0 e^{\mu_1 x}\\
			\mathit{K}(x) = \mathit{K}_0 
		\end{array}\right., \label{eq:lossy1} \\
		& \mathrm{with \;}  Z(L)=\frac{p}{v}\Big|_{x=L} = \infty, \label{eq:lossy2}
	\end{align}
	as shown in Fig.~(\ref{fig:4}b). The proposed MTMM model is actually applicable to lossy EMs as well, but it requires complex $\tilde{\rho}_0$, whose dispersion should be determined by the concrete structures in real systems. For the subwavelength acoustic dipoles based on perforated plates that we will realize in the experiments, the dispersive material properties should adopt
	
	\begin{align}
		\tilde{\rho}_0 = \rho_0\chi(\omega) = \rho_0 \left( 1+i\sqrt{\frac{\beta}{\omega}}\right), \label{eq:dipole_para}
	\end{align}
	where $\beta $ is the dissipation factor. The derivation of Eq.~(\ref{eq:dipole_para}) is available in Supplementary Materials, Section S5. By replacing Eq.~(\ref{eq:lossless1}) by Eq.~(\ref{eq:lossy1}), we can derive the explicit form of the reflection of a lossy EM with the hard boundary [Eq.~(\ref{eq:lossy2})]
	\begin{equation} \label{eq:S11_analy_hard}
		S_{11} = \frac{i \sqrt{\chi} \mathcal{F}\left(1, 0, \tilde{\xi}_0, \tilde{\xi}_L\right) - \mathcal{F}\left(0, 0, \tilde{\xi}_0, \tilde{\xi}_L\right)}{i \sqrt{\chi} \mathcal{F}\left(1, 0, \tilde{\xi}_0, \tilde{\xi}_L\right) + \mathcal{F}\left(0, 0, \tilde{\xi}_0, \tilde{\xi}_L\right)},
	\end{equation}
	where $ \tilde{\xi}_0 = \xi_0 \sqrt{\chi}$, $ \tilde{\xi}_L = \xi_L \sqrt{\chi}$, and $\mathcal{F}\left(v_1, v_2, x_1, x_2\right)$ follows the same definition of Table (\ref{tab:2}).
	Since the dissipation is inevitable in real sample, our motivation is to consider a lossy EM, whose $\beta$ can be designed to make its reflection close to that of the corresponding lossless EM, with a leaky backing. It turns out that if {\color{black}$\beta = 630\,\mathrm{rad/s}$}, the approximation is observed to be valid when $\omega L/c_0 >0.4$ [see Fig.~(\ref{fig:4}e)]. It is displayed in Figs.~(\ref{fig:4}c) and (\ref{fig:4}d) that the impedance spectra are also similar in the two cases, except the low frequency behaviors. The above results indicate that a finite lossy EM can mimic the reflection behavior of an ideal lossless EM with leaky backing boundary, laying the foundation for our experimental realization of anti-reflection meta-layer. It should be emphasized, however, that the suitable dissipation $\beta$ of a lossy EM actually depends on the impedance contrast, given by Eq.~(\ref{eq:ratio}).}
	
	\begin{figure*}[t!]
		\centering
		\includegraphics[width=0.99\textwidth]{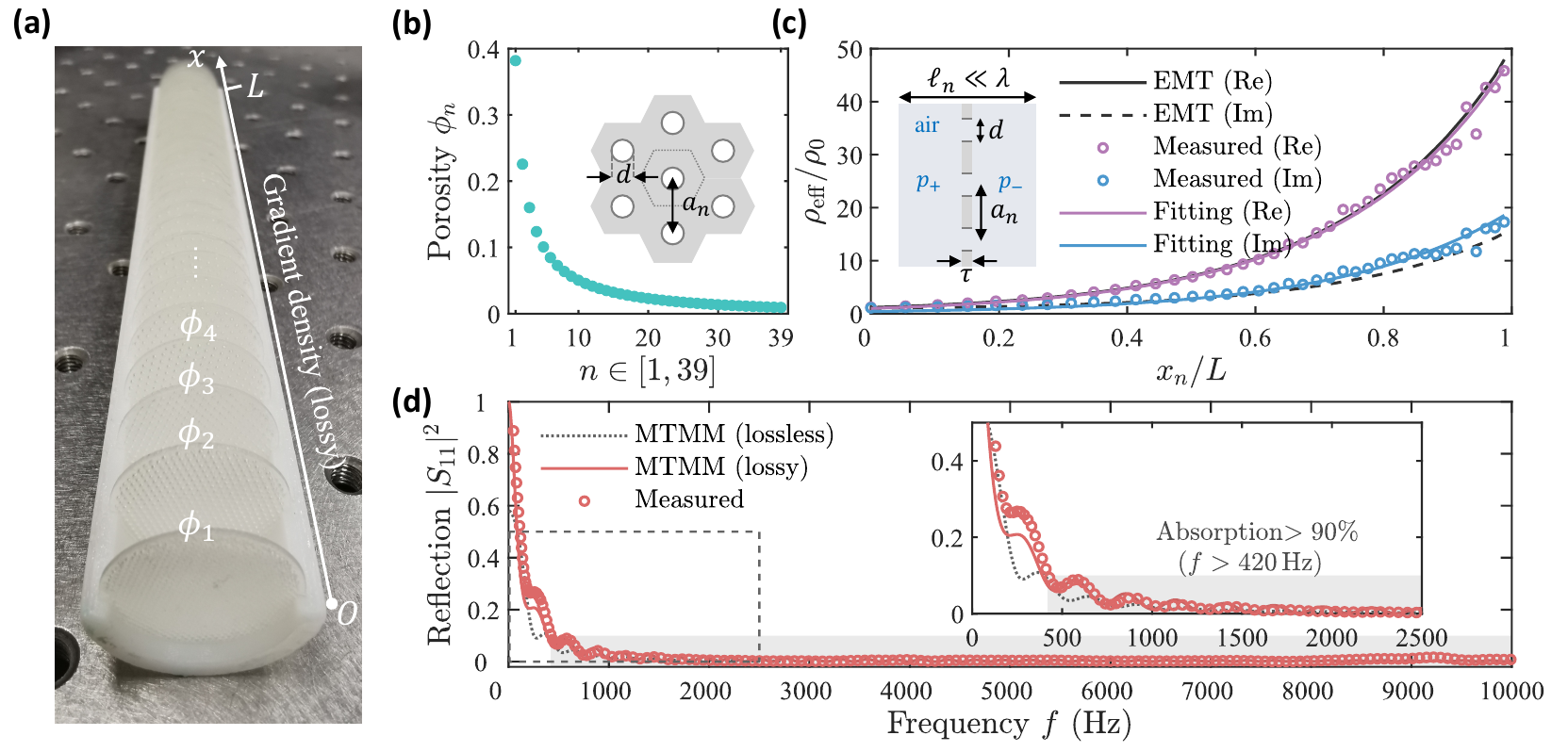}
		\caption{\label{fig:5} (a) The photo of the fabricated sample, with its cover removed for better illustration. The pore spacing is larger as the label $n$ of $\phi_n$ increases. (b) The porosity distribution generated by the proposed design scheme. The inset depicts the front view of the perforated plate. (c) The measured effective density of individual unit cells at $1000\,\mathrm{Hz}$, plotted as function of their location $x_n$. In addition, the obtained fitting function is $(1+ 0.04 i)\exp( 19.4\,\mathrm{[m^{-1}]} x)$. The inset is the side view of a unit cell. (d) The measured reflection of the overall assembled sample. The gray region from 420 Hz to 10000 Hz is determined by $ 10 \% $ reflection (or $ 90 \% $ absorption) as the threshold. Theoretical MTMM-based results are taken from Eq.~(\ref{eq:s11_mtmm}) and Eq.~(\ref{eq:S11_analy_hard}), respectively. The inset is an enlarged view of the low frequency band.}
	\end{figure*}
	
	\section{Experimental realization of anti-reflection meta-layer}
	
	\subsection{Acoustic dipoles with tunable density}
	To realize the lossy EM in Fig.~(\ref{fig:4}b) with excellent anti-reflection property, we seek real structures to realize these predetermined density properties. Therefore, the candidate we found is the perforated plate \cite{maa1998potential}, which is a type of acoustic dipole in airborne acoustic systems. The geometric parameters of a perforated plate include its thickness $\tau$, hole diameter $d$, and the spacing $a$ between holes. We also specify that the unit of the acoustic dipole is composed of air cavity with a length of $\ell$. In addition, the porosity is the perforated area ratio $ \phi = \pi d ^2/ (2 \sqrt{3} a^2) $ for hexagonal pore distribution. According to effective medium theory (EMT) \cite{cobo2019multiple,yang2014homogenization}, in the non-resonant band, the effective bulk modulus of a perforated plate is the same as that of air\footnote{From now on, $\rho_0 $ and $\mathit{K}_0$ denote the density and bulk modulus of air, respectively.}, i.e., $\mathit{K}_\mathrm{eff}\cong \mathit{K}_0$ due to the pure dipole nature. Here, the non-resonant band refers to the frequency range where $\ell\ll\lambda$ and the relevant effective properties are approximately non-dispersive. By contrast, the effective density is what we can regulate, which writes
	\begin{align}\label{eq:rho_eff}
		\rho_\mathrm{eff} &= \frac{\rho_0}{\ell} \left[\ell-\tau+ \frac{\tau+\delta \tau}{\phi}\left(1+i \sqrt{\frac{\beta}{\omega}} \right)  \right],
	\end{align}
	where the end correction $\delta \tau = 0.85d \mathrm{F}(\phi)$ and the Fok function
	\begin{equation}\label{eq:Fok}
		\mathrm{F}(\phi) = 1-1.41\phi^{\frac{1}{2}}+0.34\phi^{\frac{3}{2}}+0.07\phi^{\frac{5}{2}}-0.02\phi^{3}+0.03\phi^{\frac{7}{2}},
	\end{equation}
	which depicts the interaction between the adjacent pores \cite{melling1973acoustic}. The first and second terms in Eq.~(\ref{eq:rho_eff}) are contributed by air cavity and pore, respectively. The thermoviscous dissipation \cite{blackstock2001fundamentals} in the pore was taken account by $\beta$. Here, we adopt $\beta =8\nu/d^2$ and the kinematic viscosity $\nu = 1.59\times 10^{-5}\,\mathrm{m^2/s}$. Derivation details of Eq.~(\ref{eq:rho_eff}) can be found in {\color{DeepBlue}Supplementary Materials}, Section S5. In our case, we obtain $d = 0.45\,\mathrm{mm}$ by solving that {\color{black}$\beta(d) = 630\,\mathrm{rad/s}$}. Here, the proper dissipation ensures the similarity between the reflection spectra in the lossy and lossless cases in Fig.~(\ref{fig:4}). By adjusting $\phi$ and $\ell$, we can manipulate the effective density over a wide range, thus acoustic dipole being tunable.
	
	
	%
	
	\subsection{Design scheme}
	By utilizing the tunable acoustic dipoles, we can now construct an equivalent EM with gradient effective density. Our idea is to stack unit cells of different acoustic dipoles to create a meta-layer with a gradually varying effective property, in order to experimentally verify the anti-reflection effect shown in Fig.~(\ref{fig:4}e). To inversely engineer geometric parameters of an array of perforated plates under the constraint of exponential density [see the fabricated sample photo in Fig.~(\ref{fig:5}a)], we use $ n $ to label each perforated plate and related parameters (the total number $ N =39$). For example, the length of the $n_\mathrm{th}$ unit cell is $\ell_n$, and the porosity of the $n_\mathrm{th}$ perforated plate is $\phi_n$.
	Our strategy is to let the real part of Eq.~(\ref{eq:rho_eff}) to follow exponential dependence
	
	\begin{equation}\label{eq:exp_con}
		\exp(\mu_1 x_n) = \frac{1}{\ell_n} \left[\ell_n-\tau+ \frac{\tau+\delta \tau(\phi_n)}{\phi_n}  \right],
	\end{equation}
	where $x_n$ is the coordinate of $n_\mathrm{th}$ plate and we set $ \tau = 0.7\,\mathrm{mm} $. We set $x_1 = 0$, $x_2 = (\ell_1+\ell_2)/2$, $x_n = (\ell_1+\ell_n)/2 + \sum_{m=2}^{n-1} \ell_m $ ($n\ge3$), and the total length $L = \ell_1/2+\sum_{n=2}^{N} \ell_n $. To derive the design of $\phi_n$, we can assign $\ell_n = \ell_1 e^{-\mu_1 x_n /3}$ and solve $\ell_1$ by designating the total length $L=0.2\,\mathrm{m}$, which can determine $\mu_1 = 20 \,\mathrm{m}^{-1}$ according to Eq.~(\ref{eq:ratio}). The necessity of introducing gradient $\ell_n$ lies in the fact that the wavelength will be suppressed if the sound speed becomes slower [see Figs.~(\ref{fig:first}b) and (\ref{fig:first}c)]. In this way, the subwavelength condition ($\ell_n\ll\lambda$) can be maintained broadbandly. By numerically solving Eq.~(\ref{eq:exp_con}) for the roots, we can obtain a list of the required values of $\phi_n$ [see the outcome in Fig.~(\ref{fig:5}b)]. As for the imaginary part of Eq.~(\ref{eq:rho_eff}), if $\phi\ll 1$ (this is true for most plates), we can approximate Eq.~(\ref{eq:rho_eff}) as the first line of Eq.~(\ref{eq:dipole_para}). Hence, we can still use MTMM with the consideration of loss by Eq.~(\ref{eq:dipole_para}) to model the meta-layer. In summary, in the design scheme, besides the plate thickness $\tau$ and total length $L$, all other geometric parameters are reverse-engineered rather than obtained through large-scale multi-parameter optimization. The detailed geometrical parameters are listed in {\color{DeepBlue}Supplementary Materials}, Section S6. As for the length scale relation, it should be noted that $1/\mu $ (the exponential decay length) $>l_n$ (the unit size) $\gg \delta_\nu= \sqrt{2 \nu/\omega} $ (the boundary layer thickness).
	
	\subsection{Effective density characterization}
	We used 3D printing technology to fabricate individual perforated plates with predetermined $\phi_n$ (The pore spacing was determined by $ a_n=d [\pi/(2 \sqrt{3} \phi_n)]^{1/2}$). We performed experimental measurements of the effective density for each individual plate before assembling individual perforated plates. The specific implementation was based on the four-microphone method via impedance tube to measure the reflection ($S_{11}$) and transmission ($S_{21}$) spectra (with phase information) of a single perforated plate at a measured frequency of $ 1000\, \mathrm{Hz} $. Experimental details can be found in {\color{DeepBlue}Supplementary Materials}, Section S7. According to Ref.~\cite{yang2015subwavelength}, the measured effective density of the $n_\mathrm{th}$ plate is given by
	
	\begin{equation}\label{eq:eff_den_exp}
		\rho_\mathrm{eff}(\omega)=i\frac{2 Z_0}{\omega \ell_n}\frac{1+(S_{11}-S_{21})}{1-(S_{11}-S_{21})}.
	\end{equation}
	It is shown in Fig.~(\ref{fig:5}c) that the measured effective density (circles) indeed follows exponential spatial dependence. The black lines are the results predicted by EMT [see Eq.~(\ref{eq:rho_eff})]. We also fitted the measured real and imaginary parts of effective density by using exponential functions (purple and blue lines), which agree well with those by EMT. The above characterization experiment demonstrates the effectiveness of our design scheme.
	
	\subsection{Meta-layer preparation}
	To combine the individual fabricated perforated plates, we also printed two corresponding shell covers. The interior of the covers has designed grooves located at $x_n$ for the installation of perforated plates. We used acoustic plasticine to seal possible gaps and assembled all components to form the overall meta-layer. The bottom of the meta-layer is a hard wall [corresponding to the top of Fig.~(\ref{fig:5}a)], and the incident port of the sound wave is the bottom of Fig.~(\ref{fig:5}a). The diameter of the internal cavity of the sample is $ 2 \,\mathrm{cm} $, which is aligned with the inner diameter of the circular impedance tube. {\color{black}The corresponding cut-off frequency of the impedance tube is $10057\,\mathrm{Hz}$, which is enough to cover the range up to $10000\,\mathrm{Hz}$, the upper limit of our measurement.} To assemble the individual perforated plates, we printed two corresponding shell covers with designed grooves located at $x_n$. We used acoustic adhesive to seal possible gaps and assembled all components to form the overall meta-layer. The bottom of the meta-layer corresponds to the top of Fig.~(\ref{fig:5}a), and the incident port of the sound wave is at the bottom of Fig.~(\ref{fig:5}a).
	
	\subsection{Reflection measurement}
	For the measurement of reflection, we used two-microphone method to evaluate the anti-reflection performance of the meta-layer. To cover such a broadband range, our experiments were carried out in two rounds (implementation details and equipment specifications available in {\color{DeepBlue}Supplementary Materials}, Section S7). The experimental results show that the meta-layer possesses the low reflection of less than $ 10\% $ from 420 Hz to 10000 Hz. Within this range, the averaged reflection $\int_{f_1}^{f_2} \frac{|S_{11}(f)|^2}{(f_2-f_1)} df =0.86\%$, where $f_1=420\,\mathrm{Hz}$ and $f_2=10000\,\mathrm{Hz}$. The reflection tends to zero as the frequency increases, which is consistent with our theoretical predictions [as shown in Fig.~(\ref{fig:5}d) comparing the theoretical (red line) and experimental (red circles) data]. {\color{black}This low reflection is corresponding to a high average absorption coefficient of $99.14\%$.} The theoretical results presented here used dimensional frequency as the horizontal axis, which are actually the same data from Fig.~(\ref{fig:4}e). Because the end of the sample we designed is a hard wall, all incident energy except for the reflection will be absorbed. Compared with other competitive acoustic absorber/anti-reflection coatings, e.g., porous materials \cite{feng2021gradient}, absorbing metamaterials \cite{yang2017optimal,zhang2021broadband,zhou2022broadband}, our meta-layer has the largest relative bandwidth, i.e., $B_w = 2(f_2-f_1)/(f_2+f_1) = 1.84$, close to its upper limit 2.
	
	\section{Concluding remarks}
	{\color{black}Thus far, our results have shown an efficient and easily implementable way to achieve an EM-based anti-reflection meta-layer by manipulating the structure of multi-layer perforated plates. Other examples suitable for demonstrating more general acoustic EMs include temperature gradient pipes \cite{musielak2006method,sujith1995exact} and solid-fluid composites \cite{cervera2001refractive, krokhin2003speed, mei2006effective}. {\color{black}. We distinguish our work with acoustic black holes \cite{zhang2021broadband,pelat2020acoustic,mi2021wave}, due to the use of ideal reflectionless eigenmodes inside EMs. The precise design of the dissipation and exponential non-uniformity is also unique for the final excellent anti-reflection performance. Furthermore, our acoustic theory can be easily extended to the case of electromagnetic waves by the following mapping
	
	\begin{equation}\label{eq:mapping}
		\left\{\begin{array}{l}
			\rho \to \epsilon \\
			\mathit{K} \to \mu^{-1}
		\end{array}\right.,
	\end{equation}
	where $ \epsilon ,\, \mu $ are the permittivity and permeability, respectively.} The detailed derivations for the electromagnetic extension are available in the {\color{DeepBlue}Supplementary Materials}, Section S8. Hence, the proposed MTMM-based design scheme, together with the excellent anti-reflection effect of EM, can have even broader impacts in microwave and optical metamaterials \cite{moore1980gradient,qu2022microwave,eleftheriades2009transmission}.}


	This article focuses on one-dimensional exponential material properties. Other types of functions, such as Pöschl–Teller function ($\mathrm{sech}^2$ type) \cite{lekner2007reflectionless,thekkekara2014optical} and power series \cite{musielak2006method,soley2023experimentally} were also examined. Additionally, Ref.~\cite{erickson2018variational} proposed a variational approach to yield optimal impedance profiles, and Ref.~\cite{dhia2018trapped} systematically investigated the frequencies where waves can go around an obstacle. It was also emphasized in Refs. \cite{krapez2016heat, krapez2017sequences, krapez2018multipurpose} that infinite number of solvable impedance profiles can be found with the aid of Liouville and Darboux transformation. However, achieving broadband anti-reflection even for one-dimensional problems is not easy. Universal broadband impedance matching design were proposed in Refs.~\cite{kim2013perfect,im2018universal}. Furthermore, Refs.~\cite{horsley2015spatial,ye2017observation} demonstrated that if real and imaginary parts of the material parameters are associated by a spatial Kramer-Kronig relation, omnidirectional anti-reflection effects can be ensured with only passive components. For cases beyond one dimension, more general theories on reflectionless modes can include aspects in multi-mode problems \cite{sweeney2020theory,sol2023reflectionless}, reciprocity constraints \cite{guo2022reciprocity}, and disordered media \cite{horodynski2022anti,stone2020reflectionless,ferise2022exceptional}. The development of transformation acoustics/optics \cite{chen2010acoustic,chen2010transformation} can also be referenced for 2D and even 3D cases. However, achieving broadband effective properties in experiments is often challenging, and the required material properties can be anisotropic \cite{chen2010acoustic}.
	
	In conclusion, by taking advantage of the specific case of EMs, our work provides a foundational complement to previous studies from the analytical perspective. Under the condition of one-dimensional exponential material properties, we utilize generalized plane-wave eigenmodes as an effective theoretical analysis tool, to enhance our understanding of the impedance matching mechanism of traditional gradient materials. Simultaneously, we simplify the design procedure with the updated model, achieving unprecedented broadband anti-reflection performance. Looking ahead, EM can also be integrated as a key component into other anti-reflection devices. We expect that our acoustic-EM-based theoretical and experimental research paradigm can be extended to higher dimensions and other wave systems to generate profound impacts.

	
	\section*{Declaration of competing interest}
	The authors declare that they have no known competing financial interests or personal relationships that could have appeared to influence the work reported in this paper.
	
	\section*{Data availability}
	The data that support the findings of this study are available upon request from the corresponding authors.
	
	\section*{CRediT authorship contribution statement}
	\textbf{Sichao Qu:} Writing – original draft, Visualization, Methodology,
	Investigation, Formal analysis, Software, Conceptualization. \textbf{Min Yang:} Writing – review \& editing, Formal analysis, Funding acquisition, Project administration, Conceptualization. \textbf{Touryu Wu:} Validation, Resources, Investigation, Data curation. \textbf{Yunfei Xu:} Validation, Resources, Investigation, Data curation. \textbf{Nicholas Fang:} Writing - review \& editing, Formal analysis, Funding acquisition, Supervision, Project administration. \textbf{Shuyu Chen:} Project administration, Funding acquisition, Resources.
	
	\section*{Acknowledgements}
	S.Q. and M.Y. wish to thank Professor Ping Sheng for useful discussion on the design of multilayer perforated plates. M. Y. and S.C. acknowledges the Green Technology Fund (Grant number GTF202110282) from Environment and Ecology Bureau of Hong Kong Government for the funding support. N. F. acknowledges the startup support (Grant number GSP181), provided by Jockey Club Charities Trust STEM Lab of Scalable and Sustainable Photonic Manufacturing.
	
	\section*{Appendix A. Supplementary materials}
	Supplementary materials to this article can be found online at xxx.
	
	\bibliography{Reference.bib}
	\bibliographystyle{elsarticle-num-names}
	
	
		
		
		
\end{document}